# Computing Masses and Surface Tension from Effective Transfer Matrices

M. Hasenbusch[a], K. Rummukainen[b] and K. Pinn[c]

[a]CERN, Theory Division, CH-1211 Geneva 23, Switzerland

[b]Indiana University, Department of Physics, Swain Hall-West 117, Bloomington, IN 47405, USA

[c]Institut für Theoretische Physik I, Universität Münster, D-48149 Münster, Germany

We propose an effective transfer-matrix method that allows a measurement of tunnelling correlation lengths that are orders of magnitude larger than the lattice extension. Combining this method with a particularly efficient implementation of the multimagnetical algorithm we were able to determine the interface tension of the 3D Ising model close to criticality with a relative error of less than 1%.

## 1. Introduction

During the last two years there has been considerable progress in the Monte Carlo simulation of interfaces separating two phases of a spin model or finite temperature QCD. There are three major methods to determine the interface tension:

1. Following Binder, by comparing the height of the maximum and minimum in the order parameter distribution [1].

2. By measuring the tunnelling correlation length $\xi_{\text{tunnel}}$ of a cylindrical system [2]. In 3 D, $\xi_{\text{tunnel}}$ is related with the surface tension $\sigma$ via

$$\xi_{\text{tunnel}} \propto \exp(\sigma L^2), \qquad (1)$$

where $L$ is the extension of the lattice in spatial direction.

3. By forcing an interface into the system by applying suitable boundary conditions [3].

A major drawback of method (2.) is the rapid increase of $\xi_{\text{tunnel}}$ with the area $L^2$. Grossmann and Laursen [4] came to the conclusion that the range of $L$ values accessible with standard techniques is not sufficient to control systematic errors due to sub-leading corrections to eq. (1).

In this talk we present a new method that allows overcoming this severe problem. Using the new method one can accurately measure tunnelling correlations length $\xi_{\text{tunnel}}$ that are several orders of magnitude larger than the extension of the lattice in time direction.

More than a decade ago several authors proposed to improve the measurement of glueball masses by considering a large number $N_{\text{op}}$ of operators and their cross-correlations [5]. Lüscher and Wolff [6] showed that in the limit of infinite separation the eigenvalues of the correlation matrix give the exact masses of $N_{\text{op}} - 1$ states. However, these results can only be applied when the lattice extension is much smaller than the largest correlation length.

Motivated by studies that describe a system with cylindrical geometry as an effective 1D model [7], we consider the order parameter on a single time-slice as an effective spin. Assuming that one can neglect couplings of effective spins with distances larger than 1 we arrive at the expression

$$T_{MN} = \sqrt{\langle \delta(m_1, M) \delta(m_{t/2}, N) \rangle} \qquad (2)$$

for the effective transfer matrix. Here we reduced the whole lattice to two effective sites. The distance of these sites is half of the extension of the lattice in the time direction $t$. The operator $\delta(m_i, M)$ is equal 1 if the order parameter $m_i$ (for the Ising model, the magnetization) of the time slice $i$ takes the value $M$. Otherwise $\delta(m_i, M)$ takes the value 0.

The effective correlation length $\xi_{\text{eff}}$ can now be computed from the eigenvalues $\lambda_{\text{eff},i}$ of the effective transfer-matrix $T_{MN}$,

$$\xi_{\text{eff},i} = -\frac{t}{2}\frac{1}{\ln(\lambda_{\text{eff},i}/\lambda_{\text{eff},0})} \qquad (3)$$

where the factor $t/2$ is due to the fact that the lattice spacing of the effective model is $t/2$ while that of the original model is 1.

In general the transformation to the effective model will lead to an action that has more than nearest neighbour couplings. Hence one can only expect that $\xi_{\text{eff},i}$ converges to $\xi_i$ in the limit $t \to \infty$. An analysis of a effective two-state system shows that for $\xi_{\text{bulk}} \ll t \ll \xi_{\text{tunnel}}$

$$\xi_{\text{tunnel}} - \xi_{\text{eff,tunnel}} \propto t^{-1}. \qquad (4)$$

We obtain the expectation values of $\delta(m_i, M)$ from a Monte Carlo simulation of the original model. In order to properly measure the effective transfer matrix, we need a good statistical coverage of all the relevant magnetizations. In order to fight the *supercritical slowing down* due to exponentially suppressed tunnelling rates, we employed a multimagnetical algorithm: instead of using the canonical probability distribution, we simulated a hand-tuned distribution which explicitly enhances the probability of the states with interfaces:

$$p(\bar{\sigma}) \propto e^{-\beta H(\bar{\sigma})} G(M_{\bar{\sigma}}), \qquad (5)$$

where $M_{\bar{\sigma}}$ is the magnetization of the configuration $\bar{\sigma}$. The function $G$ is tuned so that the magnetization probability $p(M) \propto \sum_{\bar{\sigma}} p(\bar{\sigma}) \delta_{M,M_{\bar{\sigma}}}$ is roughly constant. From the measured $p(M)$ we can recover the canonical distribution

$$p_{\text{can}}(M) \propto p(M)/G(M). \qquad (6)$$

The update was implemented with a demon algorithm, which enabled us to use very efficient multi-spin coding. For details, we refer to [8–10].

## 2. Monte Carlo Results for the 2D Ising Model

We did simulations in the broken phase, at $\beta = 0.47$. This value is low enough for the tunnelling correlation length to become very large, even with modest $L$, but is close enough to $\beta_c$ for the bulk correlation length $(4.349\ldots$ when $L = \infty)$ to be still substantially larger than 1. We performed simulations for lattices with spatial extensions $L = 16, 32$ and $64$, and time-like extensions $t = L/2, L$ and $2L$. The statistics of the runs was typically $5 \times 10^7$ sweeps. The results are summarized in table 1. We find an impressive reproduction of the tunnelling correlation length that gets as large as 44014 on the $L = 64$ lattice. The convergence of $\xi_{\text{eff,tunnel}}$ towards the exact result is consistent with eq. (4).

We also tried to reproduce the large tunnelling correlation length on the $L = 64$ lattice using the standard technique of fitting the correlation functions of time slice magnetizations. However, this did not lead to any sensible result.

## 3. Monte Carlo Results for the 3D Ising Model

We simulated the 3D Ising model at $\beta = 0.225$. The results for the tunnelling correlation length are summarized in table 2. The typical statistics is again $5 \times 10^7$ sweeps. In fig. 1 we show the interface tension, obtained from the correlation length $\sigma = \log(2\xi_{\text{eff,tunnel}})/L^2$, and, using the same data, from the histogram analysis. The extrapolation $1/L^2 \to 0$ gives consistent results, provided that we discard the cubical volumes from the histogram analysis. The correlation length measurement seems to have less severe finite-size effects than the histogram method. Our result for the infinite volume limit of the surface tension at $\beta = 0.225$ is $\sigma = 0.00744(3)$.

## 4. Summary and Conclusion

Using the effective transfer-matrix method we have measured the tunnelling correlation lengths up to 229000000 for the 3D Ising model. We showed that the systematic errors of the method are under control. The application of a multi-magnetical algorithm combined with an efficient demon implementation allowed us to determine the interface tension with a relative error of less than 1%.

Table 1
Estimates of $\xi_{\text{eff}}$ for 2D Ising model with $\beta = 0.47$, obtained from the effective transfer matrix.

| $L$ | $t = L/2$ | $t = L$ | $t = 2L$ | $Exact$ |
|---|---|---|---|---|
| 16 |  | 76.6(8) | 79.1(5) | 78.159... |
| 32 | 656(13) | 732(11) | 760(10) | 753.48... |
| 64 | 35400(1900) | 41000(2200) | 43500(1600) | 44014.4... |

Table 2
Same as Table 1 for 3D Ising model with $\beta = 0.225$

| $L$ | $t = L$ | $t = 2L$ | $t = 3L$ |
|---|---|---|---|
| 28 | 2553(37) | 2650(64) | 2667(79) |
| 32 | 15470(280) | 17230(490) | 15560(590) |
| 36 |  | 115800(15000) |  |
| 42 | 3540000(160000) | 3750000(240000) |  |
| 48 |  | 229000000(19000000) |  |

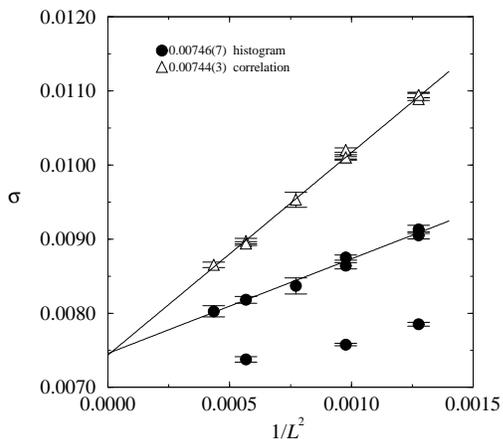

Figure 1. The interface tension measured with the correlation length and with the histogram method. The three cubical volumes have been excluded from the histogram analysis.